\documentclass[aps,prb,reprint,amsmath,amssymb,floatfix,citeautoscript]{revtex4-2}
\usepackage{graphicx}
\usepackage{color}
\usepackage{bm}
\usepackage{tabularx,booktabs}
\usepackage{multirow}
\usepackage[colorlinks,linkcolor=blue,citecolor=blue,urlcolor=blue]{hyperref}

\begin{document}

\title{Approximate Excited-State Potential Energy Surfaces for Defects in Solids}

\author{Mark E. Turiansky}
\email{mark.e.turiansky.ctr@us.navy.mil}
\affiliation{US Naval Research Laboratory, 4555 Overlook Avenue SW, Washington, DC 20375, USA}

\author{John L. Lyons}
\affiliation{US Naval Research Laboratory, 4555 Overlook Avenue SW, Washington, DC 20375, USA}

\date{\today}

\begin{abstract}
    A description of electron-phonon coupling at a defect or impurity is essential to characterizing and harnessing its functionality for a particular application.
    Electron-phonon coupling limits the amount of useful light produced by a single-photon emitter and can destroy the efficiency of optoelectronic devices by enabling defects to act as recombination centers.
    Information on atomic relaxations in the excited state of the center is needed to assess electron-phonon coupling but may be inaccessible due to failed convergence or computational expense.
    Here we develop an approximation technique to quantify electron-phonon coupling using only the forces of the excited state evaluated in the equilibrium geometry of the ground state.
    The approximations are benchmarked on well-studied defect systems, namely C$_{\rm N}$ in GaN, the nitrogen-vacancy center in diamond, and the carbon dimer in h-BN.
    We demonstrate that the zero-phonon line energy can be approximated with just a single mode, while the Huang-Rhys factor converges by including displacements up to the second nearest neighbors.
    This work also provides important insight into the success of the widely utilized one-dimensional accepting-mode approximation, specifically demonstrating that the accepting-mode Huang-Rhys factor is a strict upper bound on the full multidimensional Huang-Rhys factor.
\end{abstract}

\maketitle

\section{Introduction}

In semiconducting or insulating material, defects and impurities play essential roles in determining the properties and overall functionality of a given material~\cite{mccluskey_dopants_2018,alkauskas_tutorial:_2016,freysoldt_first-principles_2014}.
While doping enables transistors~\cite{mccluskey_dopants_2018}, recombination centers limit the efficiency of optoelectronic devices~\cite{stoneham_theory_1975,shockley_statistics_1952,hall_electron-hole_1952,alkauskas_first-principles_2014,kavanagh_rapid_2021,dou_chemical_2023}.
More recently, so-called quantum defects are being explored for applications in quantum information science~\cite{bassett_quantum_2019,wolfowicz_quantum_2021,dreyer_first-principles_2018,turiansky_rational_2024}.
Arguably the description of a defect or impurity is incomplete without a description of its coupling to the host lattice.
Not only does the lattice distort upon the introduction of the defect or impurity, but also when the defect or impurity changes its electronic configuration---a manifestation of electron-phonon coupling.
(For the rest of the paper, we will collectively refer to both defects and impurities as defects for simplicity.)
The signatures of electron-phonon coupling are apparent in almost all observable quantities of a defect.

\begin{figure}[htb!]
    \centering
    \includegraphics[width=\columnwidth,height=0.5\textheight,keepaspectratio]{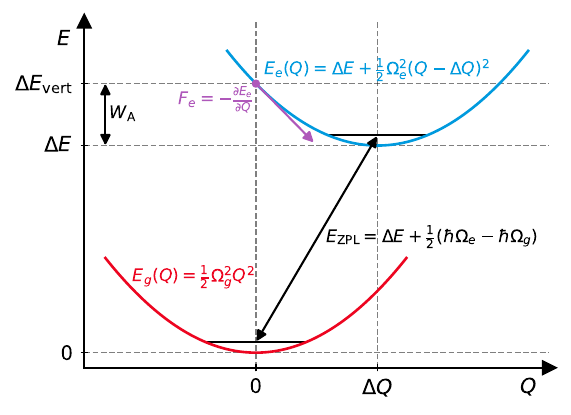}
    \caption{\label{fig:ccd}
        A schematic configuration-coordinate diagram.
        The ground- (red) and excited-state (blue) potential energy surfaces are shown in the harmonic approximation, with their dependence on the accepting mode $Q$ given.
        The vertical excitation energy $\Delta E_{\rm vert}$, energy difference between the minima $\Delta E$, relaxation energy $W_{\rm A}$, and zero-phonon line energy $E_{\rm ZPL}$ are labeled.
        $\Delta Q$ is the mass-weighted difference in atomic geometries, describing the lattice relaxation between the two equilibrium configurations.
        The relevant forces in the excited state $F_e = - \partial E_e (Q = 0) / \partial Q$ are labeled in purple.
        Replacing the accepting mode with the force mode $Q_{\rm F}$ as the configuration coordinate produces a similar picture and represents the first level of approximation used in this work.
    }
\end{figure}

The role of electron-phonon during an electronic transition at a defect can be understood using a configuration-coordinate diagram (CCD)~\cite{stoneham_theory_1975,stoneham_non-radiative_1981}, shown in Fig.~\ref{fig:ccd}.
In a CCD, there is a ground state, whose equilibrium atomic configuration is set to the origin, and an electronic excited state.
The excited state could be a different charge state of the defect (with an implicit carrier in the bulk bands), as is usual in charge trapping, or it could be an internal excited state of the defect corresponding to a different occupation of the defect orbitals, which is a typical scenario for quantum defects.
Both cases will be considered in the present investigation.
As a result of electron-phonon coupling, the atomic geometry of the excited state differs from that of the ground state.
The vector pointing along the difference in atomic geometries allows us to uniquely define a vibrational mode, known as the accepting mode~\cite{stoneham_theory_1975,stoneham_non-radiative_1981}.
All of the atomic relaxation is thus captured by the accepting mode, and this allows us to map the multidimensional relaxation onto a simple 1D picture represented by the CCD.

This mapping is not just a qualitative tool: it is the basis of rigorous first-principles calculations.
Indeed first-principles calculations are an indispensable tool for describing defects and impurities~\cite{freysoldt_first-principles_2014,dreyer_first-principles_2018,seo_first-principles_2024} and their coupling to the lattice~\cite{zhang_first-principles_2024}.
Methodologies exist to explore the various incarnations of electron-phonon coupling, from the intricate details of the phonon sideband~\cite{alkauskas_first-principles_2014-1,razinkovas_vibrational_2021} to nonradiative decay rates~\cite{alkauskas_first-principles_2014,turiansky_nonrad_2021,shi_ab_2012,nichols_multiphonon_2025}.
In the case of nonradiative transitions, the so-called accepting-mode approximation, in which this single phonon mode is used to approximate the full multidimensional system, has been employed successfully~\cite{alkauskas_first-principles_2014,shi_comparative_2015,wickramaratne_comment_2018}.
The accepting-mode approximation is also effective for calculating luminescence spectra when electron-phonon coupling is strong~\cite{stoneham_theory_1975,alkauskas_first-principles_2012,alkauskas_tutorial:_2016}.

These methodologies rely on the ability to describe the change in geometry upon change in electronic configuration.
This information, however, may be inaccessible for a variety of reasons that are usually related to numerical convergence or computational expense.
Consider the example of an internal transition at a quantum defect.
The ground state of the system is generally easily accessible using standard techniques within density functional theory (DFT).
On the other hand, the description of the excited state is non-trivial, as DFT is a ground-state theory.
One approach is constrained-occupation DFT, which---while lacking formal justification---works remarkably well for some excited states~\cite{jones_density_1989,mackoit-sinkeviciene_carbon_2019,gali_ab_2019,czelej_transition-metal-related_2024}.
It can, however, fail dramatically by never converging for a different excited state~\cite{xiong_scf_2025,*[{See the appendix of }][] bulancea-lindvall_temperature_2024,reimers_understanding_2018}, even those of the very same defect.
Another approach is the use of higher levels of theory such as time-dependent DFT~\cite{jin_excited_2023,jin_vibrationally_2022} or wavefunction-based methods through embedding techniques~\cite{muechler_quantum_2022,ma_quantum_2021,chen_advances_2025}, which are formulations specifically designed to address excited states.
They are also significantly more expensive, and it is no surprise that atomic relaxation is usually neglected when these theories are applied.
More generally, the Materials Genome Initiative~\cite{yan_case_2024} has sparked interest in big-data, high-throughput approaches to defects~\cite{goyal_computational_2017,mosquera-lois_machine-learning_2024,kavanagh_identifying_2025,thomas_substitutional_2024,kumagai_insights_2021}, which necessitate minimizing computational cost even at the DFT level.

In this work, we present a technique to approximate the role of phonons during electronic transitions, using minimal information about the excited state.
First we define a mode parallel to the forces in the excited state when evaluated in the ground-state equilibrium geometry, inspired by the accepting-mode approximation.
This enables an estimation of the relaxation energy during the transition and therefore the zero-phonon line (ZPL) energy.
However, to resolve the coupling to phonons, we include additional modes in the model, calculated using an ``equal-mode'' approximation.
We demonstrate that the strength of electron-phonon coupling converges rapidly as a function of included displacements near the defect.
Based on this multidimensional treatment, we provide insights into the accepting-mode approximation.

The paper is outlined as follows.
In Sec.~\ref{sec:assume}, we outline the basic assumptions that underlie the approach.
We first investigate a model with the mode aligned with the forces in Sec.~\ref{sec:fm} and generalize to a multidimensional model in Sec.~\ref{sec:md}.
In Sec.~\ref{sec:pl}, we provide an example application of this approach for the calculation of luminescence spectra.
Based on the multidimensional approach, insights into the accepting-mode approximation are discussed in Sec.~\ref{sec:am}.

\section{Benchmarks and Theoretical Basis}
\label{sec:assume}
To assess the usefulness of the methodology to follow, we will perform first-principles calculations on a substitutional carbon impurity (C$_{\rm N}$) in GaN~\cite{lyons_carbon_2010,lyons_effects_2014,alkauskas_first-principles_2014}, the carbon dimer (C$_{\rm B}$-C$_{\rm N}$) in h-BN~\cite{mackoit-sinkeviciene_carbon_2019}, and the nitrogen-vacancy (NV) center in diamond~\cite{gali_ab_2019,alkauskas_first-principles_2014-1}.
These systems are chosen as they are technologically relevant, cover polar and non-polar materials, include a two-dimensional material, and have varying strengths of electron-phonon coupling.
Moreover, these systems do not have trouble with excited-state relaxations~\cite{lyons_carbon_2010,mackoit-sinkeviciene_carbon_2019,gali_ab_2019}, allowing us to explicitly compute the desired quantities and compare with the proposed methodology.
The atomic geometries and a schematic depiction of their associated Kohn-Sham states are shown in Fig.~\ref{fig:atoms}.
Details of the DFT calculations are described in Appendix~\ref{app:comp_det}.

\begin{figure}
    \centering
    \includegraphics[width=\columnwidth,height=0.7\textheight,keepaspectratio]{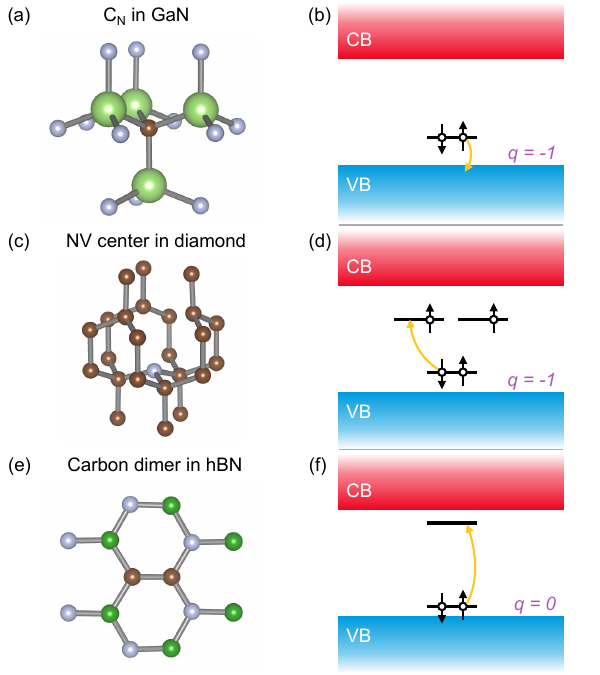}
    \caption{\label{fig:atoms}
        Atomic geometries (left column) and schematic Kohn-Sham diagrams (right column) for C$_{\rm N}$ in GaN [first row, (a) and (b)], the NV center in diamond [second row, (c) and (d)], and the carbon dimer in hBN [third row, (e) and (f)].
        C atoms are shown in brown, B atoms in dark green, Ga atoms in light green, and N atoms in lavender.
        Only atoms out to second nearest neighbor of the defect are shown.
        The charge state $q$ is labeled in purple for the corresponding Kohn-Sham diagram.
        The conduction band (CB) is shown in red, and the valence band (VB) in blue.
        Approximate locations of defect levels in the band gap are indicated by black lines.
        The white circles indicate electrons with the direction of the arrow signifying spin up or down.
        Yellow arrows denote the transition of interest.
    }
\end{figure}

We examine two different types of excitations: internal transitions and band-to-defect.
For the carbon dimer and the NV center, we consider an internal transition in which an electron is lifted to a higher defect orbital to construct the excited state.
In contrast, we consider a carrier capture process for C$_{\rm N}$ in GaN.
The ground state is the neutral charge state, and the excited state corresponds to a negatively charged defect with a hole in the valence band.
It follows that the transition energy for such a transition is related to the position of the charge-state transition level with respect to the band edge~\cite{alkauskas_first-principles_2014,freysoldt_first-principles_2014}.
This case demonstrates that the proposed methodology can be used to estimate charge-state transition levels.

For a given transition, we denote the atomic coordinates as $\{ {\bm R}_{e/g} \}$ for the ground $g$ or excited $e$ state.
A superscript ``$0$'' will denote the equilibrium geometry.
We assume that the equilibrium geometry $\{ {\bm R}_g^0 \}$ and vibrational modes of the ground state can be obtained.
Furthermore, we assume that the total energy $E_e (\{ {\bm R}_g^0 \})$ and atomic forces ${\bm F}_{e;I} (\{ {\bm R}_g^0 \}) = - \partial E_e (\{ {\bm R}_g^0 \}) / \partial {\bm R}_I$ in the excited state can be evaluated in the equilibrium geometry of the ground state---but no other geometry.
This assumption is useful for excited-state calculations that fail because of atomic relaxation (e.g., in constrained-occupation DFT) and for when evaluation of the forces at multiple geometries is prohibitively expensive (e.g., at higher levels of theory or in high-throughput calculations).
We define the vertical excitation energy in the equilibrium geometry of the ground state as
\begin{equation}
    \label{eq:vert}
    \Delta E_{\rm vert} = E_e (\{ {\bm R}_g^0 \}) - E_g (\{ {\bm R}_g^0 \}) \;.
\end{equation}
This quantity describes the excitation energy in the absence of atomic relaxation and corresponds to the average absorption energy when coupling to phonons is considered~\cite{stoneham_theory_1975,alkauskas_tutorial:_2016}.

Throughout this work, we will exploit the harmonic approximation, assuming that the potential energy surface depends quadratically on atomic displacements.
This assumption is valid for a wide range of defects and impurities in different materials (as exemplified by our three benchmark cases described above).
It is, however, not universally valid.
Since we rely on the harmonic approximation, we would not expect the methodology developed here to be applicable for cases where anharmonicity is pronounced.
For example, we would not expect to be able to describe the potential energy surface of \textit{DX} centers~\cite{chadi_energetics_1989}, many defects in the hybrid perovskites whose soft lattice leads to significant anharmonicity~\cite{zhang_defect_2022,zhang_minimizing_2021,whalley_giant_2021,meggiolaro_iodine_2018}, or more generally when metastability is present~\cite{kavanagh_identifying_2025}.

Our goal is to devise a model that gives us maximal information on the electron-phonon coupling in the system with the minimal number of additional calculations.
There are two quantities that encode this information:
the ZPL energy $E_{\rm ZPL}$, which is the energy for which no phonons are emitted during the transition, and the Huang-Rhys factor $S$, which quantifies the strength of electron-phonon coupling.
The Huang-Rhys factor is related to the Debye-Waller factor $e^{-S}$, which determines the relative weight of the ZPL to the phonon sideband in luminescence.
In addition to being features observable in the optical experiments, these quantities also encode information on whether nonradiative transitions are possible.

As discussed in the introduction, the accepting-mode approximation coincides with the description of the CCD (Fig.~\ref{fig:ccd}).
The relaxation energy along the accepting mode can be defined as $W_{\rm A} = \frac{1}{2} \Omega_e^2 \, \Delta Q^2$, where $\Omega_e$ is the frequency of the accepting mode in the excited state.
$\Delta Q$ is the mass-weighted difference in atomic geometries:
\begin{equation}
    \label{eq:dQ}
    \Delta Q^2 = \sum_I M_I \lvert {\bm R}_g^0 - {\bm R}_e^0 \rvert^2 \;.
\end{equation}
From the relaxation energy, the energy difference between the minima of the ground and excited states is $\Delta E = \Delta E_{\rm vert} - W_{\rm A}$.
This energy difference can be used to define the ZPL energy $E_{\rm ZPL} = \Delta E + \frac{1}{2} (\hbar\Omega_e - \hbar\Omega_g)$.
The contribution from the difference of harmonic oscillator zero-point energies (the second term) is usually minimal and typically neglected;
moreover we will assume the same frequencies in the ground and excited state, making $E_{\rm ZPL}$ and $\Delta E$ exactly equal.
While we will use these two quantities interchangeably, this distinction is worth noting.
In the accepting-mode approximation, the Huang-Rhys factor is expressed by~\footnote{
    A Huang-Rhys factor can be equivalently defined in the ground state, using $\Omega_g$, which is more relevant for luminescence experiments.
    Since we are interested in excited-state relaxation, we work with the associated quantities.
    However, we will also assume the phonon frequencies are the same, making the two definitions of the Huang-Rhys factor identical anyway.
}
\begin{equation}
    \label{eq:S_A}
    S_{\rm A} = \frac{W_A}{\hbar\Omega_e} = \frac{\Omega_e \, \Delta Q^2}{2\hbar} \;.
\end{equation}
The accepting mode clearly illustrates the importance of being able to resolve the atomic relaxation, as the strength of electron-phonon coupling is proportional to it.

\section{Force-Mode Model}
\label{sec:fm}
As a first step, we define a vibrational mode parallel to the forces in the excited state, which we will refer to as the force mode.
The coordinate associated with this vibrational mode is labeled ${\bm Q}_{\rm F}$.
This is our only direction in the multidimensional space of atomic displacements given to us by our initial assumptions.
We build an effective CCD (Fig.~\ref{fig:ccd}) and a one-dimensional model of the potential energy surfaces, given by
\begin{align}
    \label{eq:pes_fm}
    E_e (Q_{\rm F}) &= E_{\rm ZPL}^{\rm F} + \frac{1}{2} \Omega_{e;{\rm F}}^2 (Q_{\rm F} - \Delta Q_{\rm F})^2 \;, \\
    E_g (Q_{\rm F}) &= \frac{1}{2} \Omega_{g;{\rm F}}^2 \, Q_{\rm F}^2 \;,
\end{align}
where $\Omega_{g/e;{\rm F}}$ is the vibrational frequency of the ground $g$ or excited $e$ state along the force mode.
$\Delta Q_{\rm F}$ is the difference in atomic geometries resulting from relaxation along this vibrational mode.
The ZPL energy under the force-mode approximation $E_{\rm ZPL}^{\rm F} = \Delta E_{\rm vert} + W_{\rm F}$, where $W_{\rm F} = \frac{1}{2} \Omega_{e;{\rm F}}^2 \, \Delta Q_{\rm F}^2$ is the relaxation energy along the force mode.

Consistent with our basic assumptions, we are able to determine $\Omega_{g; {\rm F}}$, which can be computed via finite differences, and $\partial E_e (Q_{\rm F} = 0) / \partial Q_{\rm F}$.
We are left with the determination of $\Omega_{e; {\rm F}}$ and $\Delta Q_{\rm F}$.
In light of our inability to resolve vibrations in the excited state, we will assume that $\Omega_{e; {\rm F}} \approx \Omega_{g; {\rm F}}$.
While this ``equal-mode approximation'' is imperfect~\cite{zhou_defect_2025}, it has served as the basis of a variety of rigorous first-principles calculations of the vibrational properties of defects~\cite{alkauskas_first-principles_2014-1,razinkovas_vibrational_2021}.
The results to follow will serve as an implicit test of the quality of the equal-mode approximation.

To determine $\Delta Q_{\rm F}$, we take a derivative of Eq.~\ref{eq:pes_fm}:
\begin{align}
    \label{eq:forces_fm}
    \frac{\partial E_e}{\partial Q_{\rm F}} \bigg\rvert_{0} &= -\Omega_{e; {\rm F}}^2 \Delta Q_{\rm F} \approx -\Omega_{g; {\rm F}}^2 \Delta Q_{\rm F} \;, \\
    &= - \sqrt{\sum_{I\alpha} (F_{e;I\alpha})^2 / M_I} \;,
\end{align}
where the second line is the explicit expression for the magnitude of the forces along the force mode.
$I$ indexes the atoms and $\alpha$ the three Cartesian directions.
Finally we obtain the Huang-Rhys factor in the force-mode approximation
\begin{equation}
    \label{eq:S_fm}
     S_{\rm F} = \frac{\Omega_{g; {\rm F}} \, \Delta Q_{\rm F}^2}{2\hbar} = \frac{1}{2\hbar \, \Omega_{g; {\rm F}}} \left(\frac{\partial E_e}{\partial Q_{\rm F}}\right)^2 \;.
\end{equation}
$\Delta Q_{\rm F}$ serves as an approximation to the true $\Delta Q$, and $S_{\rm F}$ as an approximation to S$_{\rm A}$.

The resulting values of $E_{\rm ZPL}$, $\Delta Q_{\rm F}$, $\Omega_{g;\rm F}$, and $S_{\rm F}$ are shown in Table~\ref{tab:results}.
A comparison with the values using the explicitly calculated atomic relaxation (labeled as ``Accepting Mode'' in Table~\ref{tab:results}) is included.
(We remind the reader that the test cases were chosen because excited-state relaxation is possible, allowing us to obtain these values for benchmarking.)
We find that---despite its simplicity---the force mode provides a reasonable approximation for $E_{\rm ZPL}$.
On the other hand, $S_{\rm F}$ is a significant underestimation of the true Huang-Rhys factor $S$.
If we compare $\Omega_{g;\rm F}$ to the accepting-mode frequency $\Omega_e$, which is determined in the true excited state, we see that $\Omega_{g;\rm F}$ is significantly larger.
This leads to a much smaller $\Delta Q_{\rm F}$ compared to the true $\Delta Q$, and in turn, an underestimation of $S$.
These errors approximately cancel to give a reasonable approximation of the relaxation energy $W_{\rm F}$ and therefore a reasonable $E_{\rm ZPL}$.
The inability to describe the coupling to phonons is perhaps unsurprising:
if this were a realistic approximation to the true potential energy surface, then atomic relaxations would not require many steps to converge.

\begin{table*}[ht!]
    \centering
    \caption{\label{tab:results}
        The ZPL energy $E_{\rm ZPL}$ with relaxation energy $W_{\rm A}$ in parentheses, mass-weighted change in atomic geometry $\Delta Q$, effective phonon frequency $\Omega$, and Huang-Rhys factor $S_{\rm A}$.
        The total Huang-Rhys factor including coupling all phonon modes $S_{\rm tot}$ is shown in parentheses for the differently determined atomic relaxations.
        Utilizing the true, explicitly calculated atomic relaxation corresponds to the row labeled as ``Accepting Mode''.
    }
    \begin{ruledtabular}
        \begin{tabularx}{\textwidth}{c c cccc} 
            Defect&Method&$E_{\rm ZPL}$ ($W_{\rm A}$) [eV]&$\Delta Q$ [amu$^{1/2}$ \AA]&$\hbar\Omega$ [meV]&$S_{\rm A}$ ($S_{\rm tot}$) \\
            \midrule
            \multirow{6}{*}{C$_{\rm N}$ in GaN}&Accepting Mode&1.06 (0.464)&1.67&37&12.4 (9.62) \\
            \cmidrule{2-6}
            &Force Mode&1.23 (0.296)&0.62&81&3.67 (3.51) \\
            &1NN&1.10 (0.421)&1.25&48&8.85 (7.84) \\
            &2NN&1.02 (0.508)&1.56&42&12.2 (11.0) \\
            &All&0.96 (0.566)&2.18&32&18.0 (14.9) \\
            &All (Excited State)&1.03 (0.489)&1.76&36&13.5 (11.8) \\
            \midrule
            \multirow{5}{*}{NV in diamond}&Accepting Mode&2.06 (0.295)&0.669&74&3.97 (3.47) \\
            \cmidrule{2-6}
            &Force Mode&2.16 (0.202)&0.342&120&1.68 (1.58) \\
            &1NN&2.13 (0.233)&0.425&104&2.25 (2.12) \\
            &2NN&2.07 (0.287)&0.558&88&3.27 (3.09) \\
            &All&2.00 (0.356)&0.794&69&5.18 (4.90) \\
            \midrule
            \multirow{5}{*}{C$_{\rm B}$-C$_{\rm N}$ in h-BN}&Accepting Mode&4.41 (0.233)&0.441&100&2.33 (1.87) \\
            \cmidrule{2-6}
            &Force Mode&4.46 (0.188)&0.203&195&0.96 (0.95) \\
            &1NN&4.44 (0.202)&0.247&166&1.22 (1.14) \\
            &2NN&4.43 (0.210)&0.274&153&1.38 (1.26) \\
            &All&4.42 (0.220)&0.424&101&2.18 (1.55) \\
        \end{tabularx}
    \end{ruledtabular}
\end{table*}

\section{Multidimensional Model}
\label{sec:md}
To build a better approximation for the coupling to phonons, we turn to a multidimensional model of the excited-state potential energy surface.
We introduce the mass-weighted atomic displacement vector ${\bm u}_{I} = \sqrt{M_I} ({\bm R}_{e;I} - {\bm R}_{e;I}^0)$, where $I$ indexes the $N$ atoms in the system.
Next we introduce the dynamical matrix,
\begin{equation}
    \label{eq:dynam_mat}
    \Phi_{I\alpha,J\beta} = \frac{1}{\sqrt{M_I M_J}} \frac{\partial^2 E_e}{\partial R_{e;I\alpha} \partial R_{e;J\beta}} \bigg\rvert_{\{ {\bm R}_e^0 \}} \;,
\end{equation}
where $\alpha$ and $\beta$ index the three Cartesian directions.
With these definitions, the potential energy surface can be expanded under the harmonic approximation as:
\begin{equation}
    \label{eq:pes_u}
    E_e (\{ {\bm u} \}) = \Delta E + \frac{1}{2} \sum_{IJ} ({\bm u}_I - \Delta {\bm u}_I) \cdot {\bm \Phi}_{IJ} \cdot ({\bm u}_J - \Delta {\bm u}_J) \;,
\end{equation}
where we have taken the ground-state energy and equilibrium geometry as the origin, and $\Delta E$ is the same energy difference between the ground and excited states that was previously defined.
With these definitions, $\Delta {\bm u}_I = {\bm R}_{e;I}^0 - {\bm R}_{g;I}^0$ is the yet to be determined displacement vector between the equilibrium geometries.
In the following, we will drop the atomic indices $I$, treating ${\bm u}$ as a $3N$-dimensional column vector and ${\bm \Phi}$ as a $3N \times 3N$ matrix.

Diagonalization of the dynamical matrix ${\bm \Phi}$ gives the normal modes of vibration:
\begin{equation}
    \label{eq:diag}
    {\bm \Phi} {\bm U} = {\bm U} {\bm \Omega}^2 \;,
\end{equation}
where ${\bm U} = ({\bm \eta}_1 \, \dots \, {\bm \eta}_{3N})$ is the matrix of eigenmodes ${\bm \eta}_i$, and ${\bm \Omega}^2 = {\rm diag}(\omega_1^2, \, \dots, \, \omega_{3N}^2)$ is the diagonal matrix of eigenvalues.
Since ${\bm \Phi}$ is a symmetric matrix, ${\bm U}$ is an orthogonal transformation matrix.
Introducing the normal mode coordinates ${\bm q} = {\bm U}^T {\bm u}$, Eq.~\ref{eq:pes_u} becomes
\begin{equation}
    \label{eq:pes_q}
    E_e(\{ {\bm q} \}) = \Delta E + \frac{1}{2} \sum_i^{3N} \omega_i^2 (q_i - \Delta q_i)^2 \;.
\end{equation}

By taking a derivative of Eq.~\ref{eq:pes_q} with respect to each normal mode, we obtain
\begin{align}
    \label{eq:dqi_3n}
    \frac{\partial E_e}{\partial q_i} \bigg\rvert_0 &= -\omega_i^2 \, \Delta q_i \;, \\
    &= - \sum_{J\beta} U_{J\beta,i} F_{e;J\beta} / \sqrt{M_J} \;,
\end{align}
which gives an expression for the $i$th normal mode's contribution to the change in geometry $\Delta q_i$.
The mass-weighted change in atomic geometries can be obtained as $\Delta Q^2 = \sum_i^{3N} \Delta q_i^2$.
We define the contribution of the $i$th mode to the relaxation energy
\begin{equation}
    \label{eq:W_3n}
    W_i = \frac{1}{2} \omega_i^2 \Delta q_i^2 \;,
\end{equation}
and the corresponding partial Huang-Rhys factor
\begin{equation}
    \label{eq:S_3n}
    S_i = \frac{W_i}{\hbar \omega_i} \;.
\end{equation}
The total relaxation energy $W_{\rm tot}$ and total Huang-Rhys factor $S_{\rm tot}$ are given by a summation over the modes for the respective terms.
Acoustic phonons and imaginary modes cannot contribute to the relaxation and are excluded from the summations.
$E_{\rm ZPL}$ is then obtained as $\Delta E_{\rm vert} - W_{\rm tot}$.

In the basis of normal modes, we can construct the accepting mode as ${\bm \eta}_{\rm A} = \Delta {\bm q} / \Delta Q$.
The accepting-mode frequency $\Omega$ is then obtained as
\begin{equation}
    \label{eq:acc_freq}
    \Omega^2 = {\bm \eta}_{\rm A} \, {\bm \Omega}^2 \, {\bm \eta}_{\rm A} = \frac{1}{\Delta Q^2} \sum_i \omega_i^2 \Delta q_i^2 = \frac{2 W_{\rm tot}}{\Delta Q^2} \;,
\end{equation}
where the definition of the total relaxation energy (as outlined above) has been used.

Once again, we employ the equal-mode approximation and compute the ground-state dynamical matrix as an approximation to the excited state.
In Table~\ref{tab:results}, we provide the results when all degrees of freedom are allowed.
Allowing for additional degrees of freedom means that the relaxation energy $W_{\rm tot}$ is strictly larger than that of the force-mode approximation.
Still, the predicted ZPL energy is reasonably close to the true value.
A significant improvement in the atomic relaxation $\Delta Q$ is obtained, resulting in an improved Huang-Rhys factor $S_{\rm A}$.

While a significant improvement, there is still a remaining discrepancy with respect to the true relaxation.
In addition to the equal-mode approximation, our description of the excited-state potential energy surface also relies on the harmonic approximation.
The harmonic approximation is most valid for small displacements from the equilibrium geometry.
Here the excited-state forces are evaluated in a geometry that may be far from the equilibrium geometry and subject to the anharmonicities of the potential energy surface.
To distinguish between the effects of these two approximations, we can perform an additional calculation using the excited-state dynamical matrix.
(We reiterate that while the excited-state dynamical matrix can be obtained for these specifically chosen test cases, it is not always possible.)
This eliminates the effect of the equal-mode approximation so that the remaining discrepancies are attributable to anharmonicity.
We perform this test for the case of C$_{\rm N}$ in GaN, which had the largest $\Delta Q$ value.
We find that the description of the atomic relaxation is indeed improved by using the excited-state dynamical matrix---but is not quite perfect:
there are anharmonicities present in the potential energy surface.

This result rationalizes the approach of Razinkovas \textit{et al.}~\cite{razinkovas_vibrational_2021}, which relies on the embedding of forces from a small supercell into that of a large supercell.
These forces, however, are not taken directly from DFT.
The atomic relaxation and vibrational modes are first calculated, and this information is used to work backwards to obtain the forces \textit{within the harmonic approximation} that would produce that relaxation.
We infer that this means there are also anharmonicities present in the potential energy surface of the NV center in diamond.
Both C$_{\rm N}$ in GaN and the NV center in diamond show no discernible anharmonicity along the accepting mode (as exemplified by their CCDs);
however, these results indicate that anharmonicity along orthogonal directions may be important.
The role of anharmonicity in nonradiative transitions has been appreciated for strongly anharmonic systems, for which even the accepting mode displays anharmonicity~\cite{zhang_defect_2022,zhang_minimizing_2021,kim_anharmonic_2019}.
Our findings suggest that anharmonicity may be important even for less obvious cases.

So far, we have shown that the force mode provides a reasonable approximation to the ZPL energy and that including all phonon modes in the system can resolve the atomic relaxation.
However, computation of all phonon modes in a defect-containing supercell is expensive, requiring $N_{\rm FD} \times 3N$ calculations where $N_{\rm FD}$ is the number of finite-difference steps (typically two).
We now address whether a low-dimensional model can be used to approximate the coupling to phonons.
Given the success of the force mode, we seek a way to generate additional modes, which are orthogonal to the force mode.
We investigate three schemes to generate additional modes:
(i) purely random modes (labeled as ``Equal''), (ii) random modes with a Gaussian envelope function centered on the defect (labeled as ``Gaussian''), and (iii) unit Cartesian displacements ordered by their distance from the defect (labeled as ``Unit'').
For all three cases, we perform a Gram-Schmidt procedure to orthogonalize the new modes to the force mode and to each other.
The second and third schemes are intended to benefit from the locality of defect relaxations.

\begin{figure}[htb]
    \centering
    \includegraphics[width=\columnwidth,height=0.5\textheight,keepaspectratio]{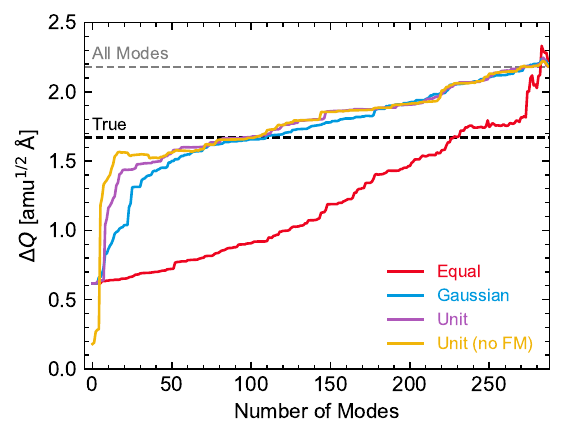}
    \caption{\label{fig:dq}
        Convergence of the mass-weighted change in atomic geometry $\Delta Q$ when the additional modes that are included are equally weighted random modes [red, scheme (i)], Gaussian weighted random modes [blue, scheme (ii)], unit displacements ordered by distance [purple, scheme (iii)], and unit displacements but without starting from the force mode (yellow).
        The dashed, horizontal lines denote the converged $\Delta Q$ including all modes (gray), which is the same for all four schemes, and the true $\Delta Q$ computed explicitly with DFT (black).
    }
\end{figure}

In Fig.~\ref{fig:dq}, we show the convergence of the atomic relaxation magnitude $\Delta Q$ as a function of the number of included modes for each scheme.
All three schemes converge to the same value when all modes are included in the model, but the rate of convergence differs.
We observe the fastest convergence for the third scheme using unit displacements orthogonalized to the force mode.
This observation raises the question of whether the force mode is needed or if only unit displacements are needed.
For this case [labeled as ``Unit (No FM)''], we find that the initial value of $\Delta Q$ is further from the converged value, but the same rapid convergence is observed.
Throughout the rest of the manuscript, we will keep the force mode and include additional modes via unit displacements [scheme (iii)].

In general, we find that within $\approx$10 additional modes, $\Delta Q$ has converged to near the full value.
This corresponds to including three displacements on the defect and its first nearest neighbors (1NN), which is intuitive as atomic relaxations tend to be localized near the defect.
Including displacements on second nearest neighbors (2NN) further improves the agreement.
The explicit $\Delta Q$ values from including modes out to 1NN and 2NN are listed in Table~\ref{tab:results}.
Since the Huang-Rhys factor $S_{\rm A}$ (Eq.~\ref{eq:S_A}), phonon frequency $\Omega$ (Eq.~\ref{eq:acc_freq}), and relaxation energy $W_{\rm A}$ depend explicitly on $\Delta Q$, these observations apply equally well to those quantities.

As noted above, including all phonon modes does not exactly reproduce the relaxation, and we attribute this to the equal-mode and harmonic approximations.
When only including modes out to 1NN or 2NN, the $\Delta Q$ values and derived quantities are closer to the true values obtained from explicit excited-state relaxations within DFT.
This suggests that there is a fortuitous cancellation:
the errors from the equal-mode and harmonic approximations appear to cancel with the errors from limiting the degrees of freedom in the system to those local to the defect.

We would summarize the essential results of the multidimensional modeling as follows:
\begin{enumerate}
    \item If only $E_{\rm ZPL}$ is required, then the force-mode approximation is sufficient.
    \item If information on the coupling to phonons (e.g., $S_{\rm A}$) or higher accuracy is required, then additional modes can be included.
    \item Additional modes can be generated by starting with the force mode and computing unit Cartesian displacements on atoms up to 1NN or 2NN from the defect, followed by an orthogonalization of the modes.
\end{enumerate}

\section{Example Application: Spectral Densities and Luminescence Lineshapes}
\label{sec:pl}
We have demonstrated that including additional modes can improve the description of atomic relaxations within our model, and therefore better describe the strength of electron-phonon coupling as quantified by the Huang-Rhys factor.
An important manifestation of electron-phonon coupling is the presence of the phonon sideband in luminescence.
The phonon sideband provides a unique fingerprint of the defect that can be compared with experiments to enable defect identification and serves as a good benchmark of the quality with which the atomic relaxation is described.
This is just one possible application of our proposed methodology to approximate atomic relaxations:
one could also examine absorption spectra~\cite{razinkovas_vibrational_2021,alkauskas_tutorial:_2016}, nonradiative transition rates~\cite{alkauskas_first-principles_2014,turiansky_nonrad_2021}, intersystem crossing~\cite{thiering_ab_2017,smart_intersystem_2021,jin_first-principles_2025}, or other phenomena that involve electron-phonon coupling.

The multidimensional formulation for the computation of the luminescence lineshape from first principles was outlined by Alkauskas \textit{et al.}~\cite{alkauskas_first-principles_2014-1}.
It is based on a generating function approach~\cite{lax_franckcondon_1952,kubo_method_1954} in which the essential quantity is the spectral density of electron-phonon coupling $S(\hbar\omega)$, given by
\begin{equation}
    \label{eq:Shw}
    S(\hbar\omega) = \sum_i S_i \, \delta (\hbar\omega - \hbar\omega_i) \;,
\end{equation}
where $S_i$ is the same partial Huang-Rhys factor from Eq.~\ref{eq:S_3n}.
In practice, the delta function in Eq.~\ref{eq:Shw} is replaced with a Gaussian whose width varies linearly with the phonon frequency (see Appendix~\ref{app:comp_det}).
One can immediately see that the derivation of the atomic relaxation within our formulation provides all the quantities necessary to evaluate the spectral density.

From the spectral density, the electron-phonon coupling spectral function is obtained as:
\begin{equation}
    \label{eq:A}
    A(\hbar\omega) = \frac{1}{2\pi} \int_{-\infty}^{\infty} dt \; e^{i\omega t - \gamma \lvert t \rvert} \, G(t) \;,
\end{equation}
where $\gamma$ is included to account for homogeneous broadening.
The generating function $G(t)$ is given by
\begin{equation}
    \label{eq:Gt}
    G(t) = \exp \left[ -i E_{\rm ZPL} t / \hbar - S_{\rm tot} + \int d(\hbar\omega) \; e^{i\omega t} S(\hbar\omega) \right] \;.
\end{equation}
Finally, the normalized luminescence intensity is obtained from
\begin{equation}
    \label{eq:luminescence}
    L(\hbar\omega) = C \, \omega^3 A(\hbar\omega) \;,
\end{equation}
where $C^{-1} = \int d(\hbar\omega) \, \omega^3 A(\hbar\omega)$ is a normalization constant.

\begin{figure*}[htb]
    \centering
    \includegraphics[width=\textwidth,height=\textheight,keepaspectratio]{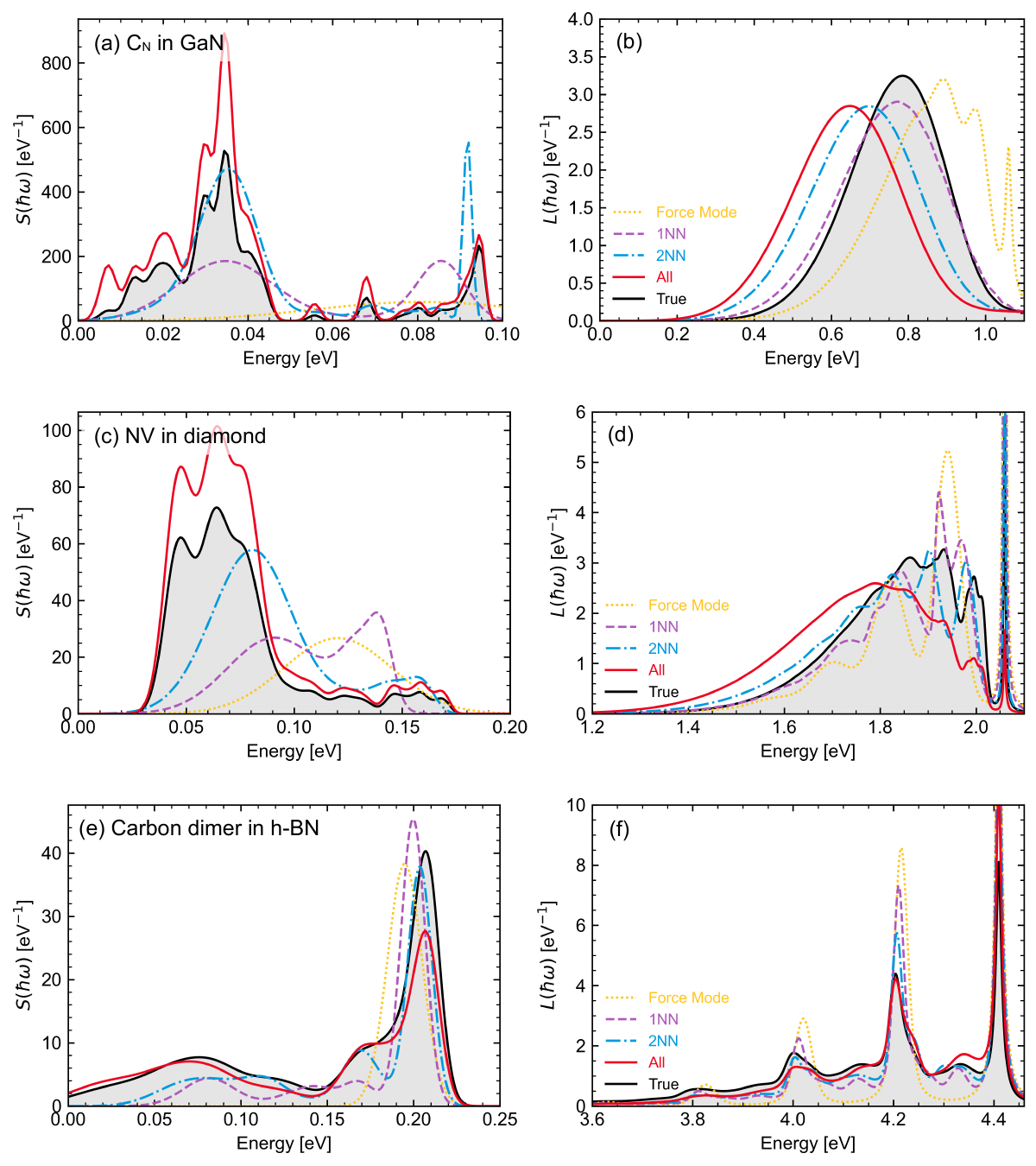}
    \caption{\label{fig:pl}
        The spectral density (left column) from Eq.~\ref{eq:Shw} and luminescence spectrum (right column) from Eq.~\ref{eq:luminescence} for C$_{\rm N}$ in GaN [first row, (a) and (b)], the NV center in diamond [second row, (c) and (d)], and the carbon dimer in h-BN [third row, (e) and (f)].
        Atomic relaxations are computed using the force-mode approximation (yellow, dotted), including displacements up to first nearest neighbor (1NN, purple, dashed), up to second nearest neighbor (2NN, blue, dashed dotted), and with all modes (red, solid).
        Results for benchmarking using the true atomic relaxation and all phonon modes are shown with black, solid lines.
    }
\end{figure*}

In Fig.~\ref{fig:pl}, we show the spectral density $S(\hbar\omega)$ and normalized luminescence intensity $L(\hbar\omega)$ for the three test cases.
The luminescence spectra are aligned to the same ZPL energy $E_{\rm ZPL}$, which is taken to be the explicitly calculated value (labeled ``Accepting Mode'' in Table~\ref{tab:results}).
It is common, when comparing with experiments, to align the ZPL to account for typical DFT errors~\cite{alkauskas_first-principles_2014-1}.
We find that the qualitative features of the spectra are well captured with these calculations, in particular when displacements up to the 2NN are included.

The spectral density for C$_{\rm N}$ in GaN [Fig.~\ref{fig:pl}(a)] rapidly converges as more displacements are included.
When all modes are included, the shape of the spectral density rather closely matches the true, explicitly calculated value;
however, we observe an overestimation of the coupling to low-energy phonon modes.
This additional coupling increases the atomic relaxation, leading to the overestimate of the true Huang-Rhys factor observed in Table~\ref{tab:results}.
Still, the Huang-Rhys factor is large for this center, and as a result, the luminescence [Fig.~\ref{fig:pl}(b)] is broad and featureless, which is captured well even with the 1NN approximation.
We note that this specific luminescence of C$_{\rm N}$ in GaN would be hard to observe experimentally, as C$_{\rm N}$ is a strong nonradiative center~\cite{alkauskas_first-principles_2014,zhao_carbon_2025} and thus the efficiency of radiative emission would be low.

Similar to C$_{\rm N}$ in GaN, the spectral density of the NV center in diamond [Fig.~\ref{fig:pl}(c)] converges rapidly and has an overestimation of the coupling to low-energy phonon modes.
Within the 2NN approximation for the relaxation, the luminescence [Fig.~\ref{fig:pl}(d)] shows qualitative agreement with the one calculated using the true atomic relaxation.
The luminescence displays strong emission into the ZPL with several phonon replicas at lower energy, albeit with a noticeable shift in the phonon replica energies.
When all phonon modes are included to predict the atomic relaxation, the positions of the phonon replicas are more accurately reproduced.
However, the overestimation of the Huang-Rhys factor leads to a shift of the average intensity to lower energies, and thus the intensity of the phonon replicas are misrepresented.
In any case, the predicted optical properties would aide in comparison with experiments, as long as the limitations of the inherent approximations are recognized.

The case of the carbon dimer in h-BN is perhaps most remarkable for its convergence.
Both the spectral density [Fig.~\ref{fig:pl}(e)] and luminescence [Fig.~\ref{fig:pl}(f)] are well described, even at the force-mode approximation.
The luminescence shows strong emission into the ZPL with phonon replicas at multiples of $\approx$200~meV from the ZPL.
In fact, the force-mode approximation performs better than the accepting mode in describing the luminescence, as the accepting mode would predict phonon replicas at multiples of 100~meV (Table~\ref{tab:results}).

The results shown in Fig.~\ref{fig:pl} constitute two approximations: (i) the number of modes included in determining the atomic relaxation, as discussed in the preceding section, and (ii) the number of modes included in evaluating the spectra.
Approximation (ii), in general, has not been thoroughly explored in the literature.
It is known that the accepting-mode approximation works well when the Huang-Rhys factor is large~\cite{alkauskas_first-principles_2012,alkauskas_tutorial:_2016}.
When the Huang-Rhys factor is moderate to small, all modes can be included~\cite{alkauskas_first-principles_2014-1}.
However, the convergence as a function of the number of included modes has not been explored, even when the true atomic relaxation is known.
In Appendix~\ref{app:conv}, we examine the effect of approximation (ii).
We find that when the approximate relaxation proposed here is used, there is little difference in the predicted luminescence intensity once differences in broadening are accounted for.
Similarly, we find that the luminescence rapidly convergences as more modes are included when the true atomic relaxation is known.

\section{Insights into the Accepting Mode}
\label{sec:am}
The multidimensional formulation allows us to gain insight into the accepting-mode approximation that has been widely used in the literature and is implicitly invoked when constructing a CCD.
The accepting mode is constructed to be parallel to the atomic relaxation during the transition.
As previously stated, this construction ensures that the relaxation energy for the accepting mode $W_{\rm A}$ is equivalent to the total relaxation energy when considering all modes $W_{\rm tot}$:
\begin{equation}
    \label{eq:acc_W}
    W_{\rm A} = \frac{1}{2} \Omega^2 \Delta Q^2 = W_{\rm tot} \;,
\end{equation}
which follows from the definition of $\Omega^2$ in Eq.~\ref{eq:acc_freq}.
This helps to rationalize the success of the accepting mode, since it exactly reproduces the relaxation energy during the transition.
As was shown in Ref.~\onlinecite{alkauskas_tutorial:_2016}, the accepting-mode approximation is most effective when electron-phonon coupling is strong (large $S$);
in this case, the peak in absorption or luminescence is the main observable and is directly related to the relaxation energy.

Implicit in this construction is the fact that $S_{\rm A}$ is generally not the same as $S_{\rm tot}$.
In fact, we can prove that $S_{\rm A}$ is an upper bound on $S_{\rm tot}$.
From Eqs.~\ref{eq:W_3n} and \ref{eq:S_3n},
\begin{equation}
    \label{eq:Stot}
    S_{\rm tot} = \frac{1}{2\hbar} \sum_i \omega_i \Delta q_i^2 \;.
\end{equation}
The summation on the right-hand side can be thought of as a dot product between a vector with elements $\omega_i \Delta q_i$ and a vector with elements $\Delta q_i$.
This allows us to apply the Cauchy-Schwarz inequality~\cite{steele_cauchy-schwarz_2004} to the right-hand side,
\begin{align}
    S_{\rm tot} &\leq \frac{1}{2\hbar} \left(\sum_i \omega_i^2 \Delta q_i^2 \right)^{1/2} \left( \sum_i \Delta q_i^2 \right)^{1/2} \;, \\
    &\leq \frac{1}{2\hbar} \sqrt{2 W_{\rm tot}} \, \Delta Q \;, \\
    &\leq \frac{1}{2\hbar} \Omega \, \Delta Q^2 \;, \\
    &\leq S_{\rm A}
\end{align}
where the definition of the accepting-mode frequency from Eq.~\ref{eq:acc_freq} has been used.
This result is explicitly verified for the cases studied here in Table~\ref{tab:results}.
We observe that $S_{\rm A}$ can overestimate $S_{\rm tot}$ by as much as 40\%, as in the case of the carbon dimer.

Being able to bound the true multidimensional Huang-Rhys factor can be a powerful tool.
For example, in the search for single-photon emitters, one wants to minimize the Huang-Rhys factor to maximize emission into the ZPL~\cite{turiansky_rational_2024}.
By having an upper-bound for the multidimensional Huang-Rhys factor, we can say \textit{at least} this much light is emitted into the ZPL.
The accepting-mode approximation, which has been the basis of a number of first-principles investigations, actually represents a more rigorous statement on the strength of electron-phonon coupling than has been previously appreciated.
Indeed, the fact that $S_{\rm A}$ in general overestimates $S_{\rm tot}$ likely also rationalizes why the accepting-mode approximation has been so successful in predicting nonradiative rates.
Since the nonradiative rate is exponentially dependent on the Huang-Rhys factor~\cite{stoneham_theory_1975,turiansky_rational_2024}, the overestimation of $S_{\rm tot}$ by $S_{\rm A}$ offsets the restricted degrees of freedom.

\section{Conclusion}
\label{sec:concl}
In summary, we have developed a novel method to address the excited-state potential energy surfaces of a given defect.
This approach enables extracting quantitative information on the coupling to phonons even when the excited-state calculation fails to converge, is too expensive to compute, or needs to be done cheaply.
We demonstrated that a model based on a single phonon mode along the forces is able to approximately capture the relaxation energy to predict the ZPL, requiring only a single additional DFT calculation in the ground state to extract the phonon frequency.
From considering a multidimensional model, we showed that the coupling to phonons, as quantified through the Huang-Rhys factor and through calculation of the luminescence spectrum, converges rapidly as additional displacements on atoms near the defect are included.
We also gained important insight into the accepting-mode approximation that is widely utilized in the literature: it constitutes an upper bound on the Huang-Rhys factor of the system.
In total, these findings provide important information on the nature of electron-phonon coupling in defect-based systems and will enable calculations of excited states that were not previously feasible.

\begin{acknowledgments}
    This work was supported by the Office of Naval Research through the Naval Research Laboratory's Basic Research Program.
\end{acknowledgments}

\section*{Data Availability}
The data that supports the findings of this study are available from the corresponding author upon reasonable request.

\bibliographystyle{apsrev4-2}

\appendix

\section{Computational Details}
\label{app:comp_det}

We perform density functional theory calculations as implemented in the VASP code~\cite{kresse_efficient_1996,kresse_efficiency_1996} version 6.4.3.
The projector augmented-wave formalism~\cite{blochl_projector_1994} is used to freeze the core electrons, and the valence electrons are represented in a plane-wave basis.
We truncate the basis at an energy of 400~eV for GaN and 520~eV for h-BN and diamond.
Ga $d$ states are included in the core.
For an accurate description of the electronic structure, we utilize the hybrid functional of Heyd, Scuseria, and Ernzerhof~\cite{heyd_hybrid_2003,heyd_erratum:_2006}.
The fraction of short-range Hartree-Fock exchange is set to 0.31 for GaN, 0.40 for h-BN, and 0.25 for diamond with the default screening parameter of 0.2~{\AA}$^{-1}$.
For h-BN, we empirically include van der Waals interactions via the Grimme-D3 scheme~\cite{grimme_consistent_2010} with the default parameters for PBE, which were also the default for HSE prior to VASP version 6.
These values are consistent with previous studies~\cite{turiansky_dangling_2019,mackoit-sinkeviciene_carbon_2019} and chosen to reproduce the experimental band gaps.

Defects are studied in a supercell configuration within periodic boundary conditions~\cite{freysoldt_first-principles_2014}.
We utilize a 96-atom supercell for GaN, which is a $3\times2\times2$ multiple of the 8-atom orthorhombic cell obtained by transforming the 4-atom primitive cell.
For h-BN, we similarly transform the primitive cell into an 8-atom orthorhombic cell and scale to a $5\times3\times2$ multiple, which contains 240 atoms.
A 216-atom supercell of diamond is constructed as a $3\times3\times3$ multiple of the conventional standard unit cell.
In all supercell calculations, the Brillouin zone is sampled with the mean-value point~\cite{baldereschi_mean-value_1973}, and the atomic coordinates are relaxed until the forces are below 5~meV/{\AA}.
The lattice parameters are kept fixed at the calculated bulk values.
Spin polarization is explicitly taken into account.

To address the excited states of the NV center in diamond and the carbon dimer in h-BN, we utilize the constrained-occupation $\Delta$SCF approach~\cite{jones_density_1989}.
The excited state of the carbon dimer is multideterminant in nature, and we correct the ZPL energy to account for this as was done in Ref.~\onlinecite{mackoit-sinkeviciene_carbon_2019}, following the suggestion of von Barth~\cite{von_barth_local-density_1979}.
However, we approximate the potential energy surface of the excited state with the potential energy surface of the mixed state, consistent with Ref.~\onlinecite{mackoit-sinkeviciene_carbon_2019}.
For the NV center in diamond, we utilize the low-symmetry electronic occupation for the excited state, which results from a Jahn-Teller distortion.
A detailed study of the luminescence properties of the NV center should account for the Jahn-Teller effect explicitly~\cite{bersuker_jahn-teller_2006,razinkovas_vibrational_2021}, but this is beyond the scope of the present study.
This choice is entirely for convenience:
the methodology proposed here is fully compatible with the fractional occupation scheme used to study the influence of the Jahn-Teller effect on the NV center's luminescence.

As noted in the main text, the delta functions in the spectral density $S(\hbar\omega)$ (Eq.~\ref{eq:Shw}) are replaced with a Gaussian.
The broadening parameter is varied linearly from $\sigma_{\rm low}$ at zero to $\sigma_{\rm high}$ at the highest phonon frequency.
In addition, there is a parameter $\gamma$ in the evaluation of the spectral function $A(\hbar\omega)$ (Eq.~\ref{eq:A}), which accounts for homogeneous broadening.
The values that were used to generate Fig.~\ref{fig:pl} are specified in Table~\ref{tab:broad}.
In Table~\ref{tab:broad}, ``True'' refers to using the explicitly calculated DFT atomic relaxation (as is the case for the rows labeled ``Accepting Mode'' in Table~\ref{tab:results}), but here all phonon modes are used to compute the luminescence, which is why the broadening parameters match the ``All'' case.

\begin{table}[ht!]
    \centering
    \caption{\label{tab:broad}
        The low $\sigma_{\rm low}$ and high $\sigma_{\rm high}$ Gaussian broadening parameters in the spectral density, as well as the homogeneous broadening parameter $\gamma$ in the spectral function.
    }
    \begin{ruledtabular}
        \begin{tabularx}{\columnwidth}{c c ccc} 
            Defect&Method&$\sigma_{\rm low}$ [meV]&$\sigma_{\rm high}$ [meV]&$\gamma$ [meV] \\
            \midrule
            \multirow{5}{*}{C$_{\rm N}$ in GaN}&True&2&1&10 \\
            \cmidrule{2-5}
            &Force&50&25&10 \\
            &1NN&15&5&10 \\
            &2NN&10&1&10 \\
            &All&2&1&10 \\
            \midrule
            \multirow{5}{*}{NV in diamond}&True&7.5&2.5&2.5 \\
            \cmidrule{2-5}
            &Force&50&25&2.5 \\
            &1NN&50&5&2.5 \\
            &2NN&30&5&2.5 \\
            &All&7.5&2.5&2.5 \\
            \midrule
            \multirow{5}{*}{C$_{\rm B}$-C$_{\rm N}$ in h-BN}&True&20&7&7.5 \\
            \cmidrule{2-5}
            &Force&10&10&7.5 \\
            &1NN&20&7&7.5 \\
            &2NN&20&7&7.5 \\
            &All&20&7&7.5 \\
        \end{tabularx}
    \end{ruledtabular}
\end{table}

\section{Number of Modes Convergence}
\label{app:conv}

\begin{figure*}[htb]
    \centering
    \includegraphics[width=\textwidth,height=\textheight,keepaspectratio]{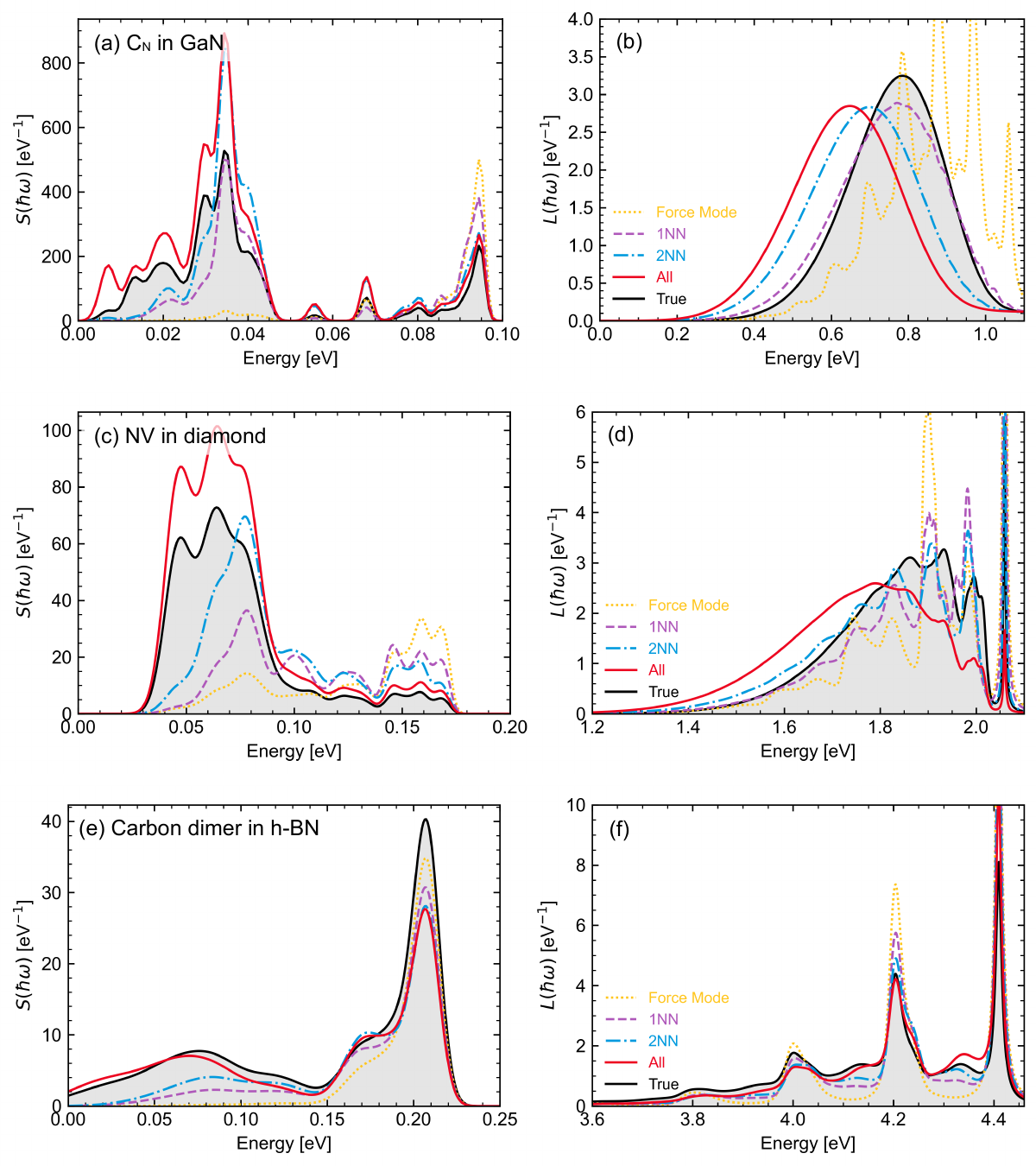}
    \caption{\label{fig:pl_all}
        The spectral density (left column) from Eq.~\ref{eq:Shw} and luminescence spectrum (right column) from Eq.~\ref{eq:luminescence} when all phonon modes are included in obtaining the spectra for C$_{\rm N}$ in GaN [first row, (a) and (b)], the NV center in diamond [second row, (c) and (d)], and the carbon dimer in h-BN [third row, (e) and (f)].
        Atomic relaxations are computed using the force-mode approximation (yellow, dotted), including displacements up to first nearest neighbor (1NN, purple, dashed), up to second nearest neighbor (2NN, blue, dashed dotted), and with all modes (red, solid).
        Results for benchmarking using the true atomic relaxation with all phonon modes are shown with black, solid lines.
    }
\end{figure*}

Within the proposed methodology, one has a choice of the number of atomic displacements included when determining the atomic relaxation.
Based on our findings, we suggested that included displacements up to second nearest neighbors of the defect provides reasonable accuracy in the coupling to phonons due to a cancellation of error.
From the obtained atomic relaxations, we produced the spectral densities and luminescence spectra in Fig.~\ref{fig:pl}.
These results, however, are also subject to a second approximation, namely the number of modes included in computing the spectra for a given atomic relaxation.
In other words, once one obtains the atomic relaxation with a fixed number of modes, the remaining modes could be calculated and included in the calculation of the spectra.
To assess the effect of this approximation, we explicitly compute the spectral density and luminescence spectrum with all phonon modes included for each atomic relaxation and each defect in our study.
The results are shown in Fig.~\ref{fig:pl_all}, noting that the labels now refer only to the number of modes used in determining the atomic relaxation and not the number of modes in the spectra.
As all phonon modes are included in the spectra, we utilize the broadening parameters specified for ``All'' in Table~\ref{tab:broad}.

As shown in Fig.~\ref{fig:pl_all}, the resulting spectral densities and luminescence intensities are qualitatively similar to those using a reduced number of modes (Fig.~\ref{fig:pl}).
The largest discrepancy is observed for the force mode, which inherits some finer structure.
Indeed, the 1NN and 2NN approximations are more qualitatively correct and inherit the finer structure of the calculation involving all modes.
Noticeably there is missing intensity in the spectral density for low energy phonon modes.
Low energy modes, such as acoustic modes, involve displacements on the entire cell, and therefore the restriction to displacements local to the defect prevents their description.
In general, the converged description of the low energy coupling requires extremely large supercells~\cite{alkauskas_first-principles_2014-1}, which is beyond the scope of this study.

\begin{figure*}[htb]
    \centering
    \includegraphics[width=\textwidth,height=\textheight,keepaspectratio]{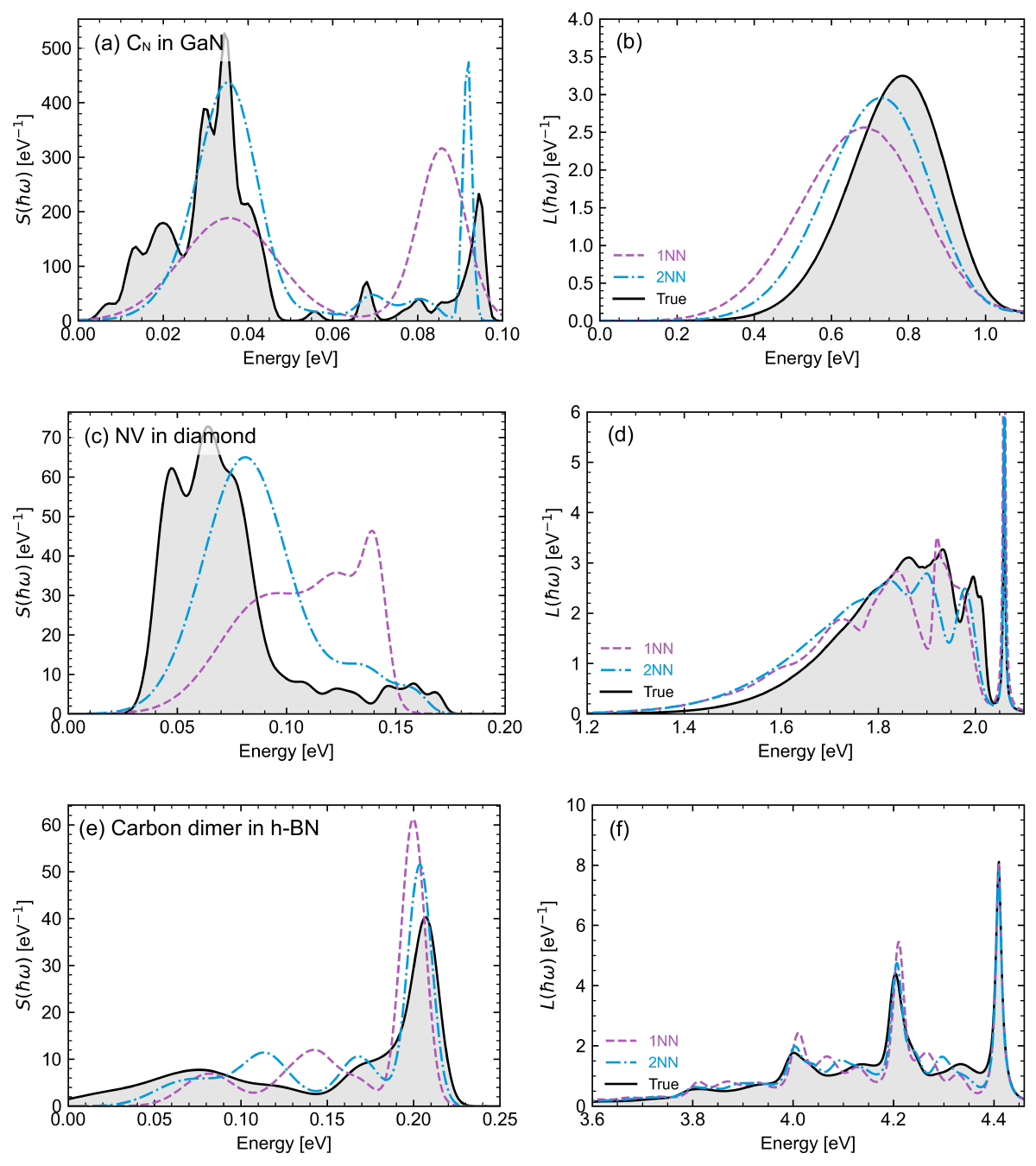}
    \caption{\label{fig:pl_full}
        The spectral density (left column) from Eq.~\ref{eq:Shw} and luminescence spectrum (right column) from Eq.~\ref{eq:luminescence} with the true atomic relaxation and a limited displacement basis for C$_{\rm N}$ in GaN [first row, (a) and (b)], the NV center in diamond [second row, (c) and (d)], and the carbon dimer in h-BN [third row, (e) and (f)].
        The basis including displacements up to first nearest neighbor (1NN, purple, dashed) and up to second nearest neighbor (2NN, blue, dashed dotted) are shown.
        Results with all phonon modes are shown with black, solid lines.
    }
\end{figure*}

We may also assess the extent of using a limited number of modes to compute the spectra when the true atomic relaxation is known.
In Fig.~\ref{fig:pl_full}, we compute the spectral densities and luminescence intensities under this restriction.
As the number of modes is restricted, the utilized broadening parameters are the same as in Table~\ref{tab:broad}.
In the case of C$_{\rm N}$ in GaN and the carbon dimer in h-BN, the predicted luminescence are qualitatively correct with the restricted basis.
For the carbon dimer, the major phonon replicas are accurately captured.
The NV center in diamond shows larger discrepancies in the finer structure, but in general, gives a reasonable approximation to the lineshape.
It may therefore be a plausible approach, when computational resources are limited and the full relaxation is accessible, to converge the calculated luminescence with respect to the shells of atoms included in the displacement basis.

\end{document}